# Superlens-Assisted Laser Nanostructuring of Long Period Optical Fiber Gratings (LPGs) for Enhanced Refractive Index Sensing

Yiduo Chen[1], Liyang Yue[1], Zengbo Wang[1,*]

[1] School of Computer Science and Electronic Engineering, Bangor University, Dean Street, Bangor, Gwynedd LL57 1UT, UK.

*Correspondence: z.wang@bangor.ac.uk

**Abstract:**

We present an innovative method to enhance Long Period Optical Fiber Gratings (LPGs) for refractive index sensing using microsphere-assisted superlens laser nanostructuring. This technique involves self-assembling a silica microsphere monolayer on LPGs' outer surface, followed by pulsed laser irradiation to generate nanoholes (300-500 nm) forming nanohole-structured LPGs (NS-LPGs). In experiments, two nanohole densities were compared for their impact on sensing performance in sucrose and glycerin solutions. The nanostructured NS-LPGs showed improved sensitivity by 16.08% and 19.57% compared to regular LPGs, with higher nanohole density yielding greater enhancement. Importantly, the permanent nanohole structures ensure durability in harsh environments, surpassing conventional surface-coating-based LPGs. Further improvements can be achieved by refining nanostructuring density and controlling nanohole size and depth. Our work represents a notable advancement in LPG sensor engineering, prioritizing surface nanostructuring over nano-coating, promising enhanced refractive index sensing applications.

## 1. Introduction

Over the past few decades, optical fiber grating sensors have witnessed significant growth and gained increasing prominence in various fields, including medicine, life sciences, security, food industry, and environmental monitoring technologies. The development of optical fiber sensing technology has been driven by its recognized advantages, such as its compact size, lightweight nature, immunity to electromagnetic interference, exceptional sensitivity, real-time monitoring capabilities, and the ability to multiplex [1-4]. Optical fiber grating sensors, including, Fiber Bragg Grating (FBG) and Long-period Grating (LPG) sensors, relies on the fundamental principle of leveraging evanescent fields created by the propagation of light within fiber-based devices and their interaction with the surrounding environment and medium [5]. While FBGs are widely used in temperature and strain sensing, LPGs are widely used in refractive index sensing and monitoring of chemical processes [6]. The LPG gratings consist of periodic changes in the fiber core's refractive index, causing coupling between the core mode and co-propagating cladding modes, leading to resonance bands in the transmission spectrum. Each resonance band corresponds to a different cladding mode and offers varying sensitivity to environmental changes [7]. The RI sensitivity of LPGs relies on the effective RI of the cladding modes, which, in turn, depends on the difference between the cladding's RI and the surrounding medium's RI. LPGs are most sensitive when the surrounding medium's RI is close to that of the cladding [8].

To enhance the RI sensitivity of LPG sensors, various strategies have been employed. These include uniform thin-film coatings such as PVA [9], PDMS [10], polyimide [11], graphene [12], graphene oxide [13], and Zn/ZnO [14], as well as nanoparticle coatings [15-17], and nanoparticle-embedded composite coatings [18, 19]. Among the different methods for depositing coatings on LPGs, the Layer-by-Layer technique is notable for its unique ability to precisely control the thickness and composition of nanoparticles in the resulting thin films. This versatile method has enabled the monitoring of a wide range of substances, including ethanol [20], ammonia [21], and low molecular analytes [22].

In this study, we introduce a new strategy to enhance the RI sensitivity of LPG sensors through direct laser nanostructuring of the LPG surface, assisted by microsphere superlenses coated on its surface, leading to the formation of nanohole-structured LPGs (NS-LPGs). The use of microsphere lenses as optical superlenses for nanoscale sub-wavelength patterning, imaging, and sensing has been an active research area since 2000 [23], and our team has been actively involved. Patterning resolutions of 80-400 nm and imaging resolutions of 45-100 nm have been reported in the literature [24-29], but no reports exist on microsphere-assisted nanostructuring of the fiber surface of LPG sensors to create NS-LPG sensors. In our experiments, we used a dip coating technique to apply a monolayer of silica microspheres to the entire surface of the optical fiber [30]. A pulsed UV laser system was then used to ablate nanoholes into the fiber's surface, with the microsphere monolayer acting as a focusing lens array, enabling the creation of numerous nanoholes. We tested the RI sensing performance of the fabricated NS-LPGs in both sucrose and glycerin solutions and compared their efficacy with standard LPG sensors. Our results showed that the NS-LPG sensor had greater sensitivity than the regular LPG sensor. We also examined the effect of nanohole density on sensing performance, finding that a higher nanohole density improves sensitivity. Following this introduction, we will provide experimental details, results and discussion, and then conclusions.

2. **Fabrication & experiment**

*2.1 Materials*

**Chemicals:** Chemical reagents were sourced from established suppliers: Potassium hydroxide (KOH, 56.11 g/mol), poly(diallyl dimethylammonium chloride) (PDDA, 200k-350k molecular weight, 20 wt% in water), high-purity sucrose ($\geq$ 99.5%), and glycerol ($\geq$ 99.5%) were all obtained from Sigma-Aldrich (UK). Deionized (DI) water used in the experiments was from Thermo Fisher Scientific Inc. (UK). Different concentration solutions were prepared with DI water, and reagents were used as received, without further purification.

**Microspheres:** Colloidal silica microspheres with a diameter of 1 micrometer, a concentration of 50 mg/ml, and a coefficient of variation (CV) less than 3% were procured from ALPHA Nanotech Inc. (United Kingdom).

**LPGs:** Long Period Fiber Gratings with a cladding diameter of 125 μm, core diameter of 8.2 μm, grating length of 20 mm, and grating period of 550 μm were sourced from oeMarket (Australia). These were fabricated from single-mode fiber (SMF-28).

*2.2 Deposition of Microsphere Monolayer*

The dip coating method was used to deposit a silica microsphere monolayer on the LPG surface using the Type G Ossila Dip Coater from Ossila, UK. The prepared solutions included 5 wt% ethanolic KOH (ethanol/water = 3:2 v/v), 10 wt% PDDA in DI water, and 10 wt% silica microspheres in DI water. The LPG was attached to the dip coater holder, rinsed with DI water, and cleaned with ethanol. It was immersed in the 5 wt% KOH solution for 20 minutes to produce a negatively charged surface. Next, it was submerged in the 10 wt% PDDA solution for 20 minutes, followed by the 10 wt% silica microspheres solution. After these steps, the LPG was air-dried at 50°C for an hour with a Fisher Scientific Heater, facilitating silica microsphere monolayer deposition [31-33] (Fig. 1a). The dip and withdrawal speeds were consistently set at 50 mm/min and 2 mm/min, respectively. The microsphere-coated LPG sample was characterized using an advanced Olympus Microscope (DSX1000) and a 3D Measuring Laser Confocal Microscope OLS5000.

*2.3 Nanohole generation*

As previously noted, microspheres can act as superlenses, focusing light to sub-wavelength spot sizes [30, 34, 35]. In our experiments, we employed a pulsed UV nanosecond laser (wavelength: 355 nm, pulse duration: 100 ns, repetition rate: 50kHz, spot size: 50 µm). This laser beam was directed onto the microsphere-coated LPG sample and scanned across using a Galvo scanner at 100 mm/s. After laser treatment, nanoholes appeared beneath each microsphere, and the majority of microspheres were removed due to the ablation process. These procedures are schematically depicted in Fig. 1(b) and Fig. 1(c). It's essential to remember that the fiber sample is cylindrical. To ensure complete surface treatment, the LPG sample needed rotation after each procedure. This was achieved with a specially designed rotational holder. It's also crucial to mention that the laser power was carefully adjusted to prevent damage to the LPG grating inside the fiber areas not covered by particles. In the experiments, two sets of nanohole densities were compared: sample **LPG1** involved a single-time microsphere deposition and laser processing, while sample **LPG2** used double-time microsphere deposition and laser processing, where microspheres were reapplied and laser-processed for a second time, resulting in a second nanohole generation. Although this process can be repeated multiple times to increase nanohole density further, our study focuses on single-time and double-time processing only. For easy comparison, we'll call the unprocessed original LPG sample **LPG0**. The nanostructured LPG samples were characterized using various imaging techniques, including dark field microscopy, scanning electron microscopy (SEM), and atomic force microscopy (AFM).

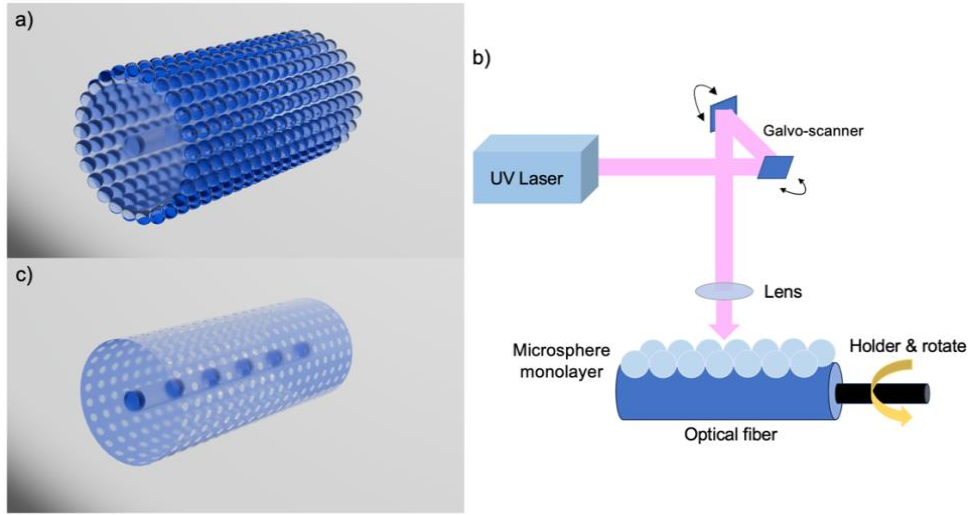

Fig. 1. Schematic of the experimental processes. a) Microsphere coating on LPG, b) Laser processing, c) Nanoholes fabricated on LPG.

*2.4 Refractive Index Sensing*

The refractive index (RI) of solutions varies with concentration, with distinct RIs for each solute [36]. LPG sensors detect RI shifts, especially around water RIs between 1.33 to 1.46, slightly below the cladding's RI. As ambient RI nears the cladding's, LPG sensitivity increases [37]. We assessed this using solutions of varying sucrose (0-70 wt%) and glycerin (0-90 wt%) concentrations.

Our setup, shown in figure 2, utilized a broadband light source (BBS, NKT Photonics SuperK COMPACT) connected to an optical spectrum analyzer (OSA, Anritsu MS9740B) via NKT Photonics SuperK Fiber Delivery on a Windows system. After each immersion, the LPG was cleaned and dried to ensure consistent readings and prevent contamination.

LPG's response is influenced by the surrounding material's RI. Changes in this RI impact the LPG's spectral output, as light attenuation wavelengths rely on the cladding mode's effective RI. The LPG transfers light from the core to specific cladding modes. Within the cladding, light diminishes due to scattering, leading to observable loss bands in the core output [38].

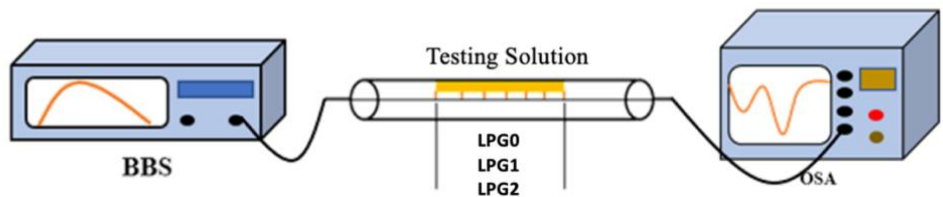

| **Sample** | **Description** | **Hole Density (hole area/total surface area)** |
|---|---|---|
| LPG0 | Original, non-processed LPG. | 0% |
| LPG1 | One-time processed LPG | 5% |
| LPG2 | Double-time processed LPG | 7.9% |

Fig. 2. Sensing testing experimental setup and samples.

## 3. Results and discussion

### 3.1 Morphology of Microsphere-coated LPG

Figure 3 presents a detailed analysis of the surface morphology of Long-Period Fiber Gratings (LPGs) coated with a monolayer of silica microspheres. Dark field optical images, as seen in Figure 3a, emphasize the efficient alignment and distribution of the microspheres across the fiber's surface. The coating procedure achieved substantial coverage, covering approximately 95% of the fiber's surface area. In Fig. 3(b), the 3D confocal image illustrates the assembly of microspheres on the curved surface of the LPG. Notably, the microspheres' monolayer exhibits uniform coverage across the entire LPG's curved surface, rather than an inconsistent distribution. Figure 3c provides a closer look at the monolayer-coated LPG, revealing silica particles with a size of 1000 nm. This magnified view clearly depicts the silica microparticles in a hexagonal honeycomb arrangement, despite defects that could exist within them.

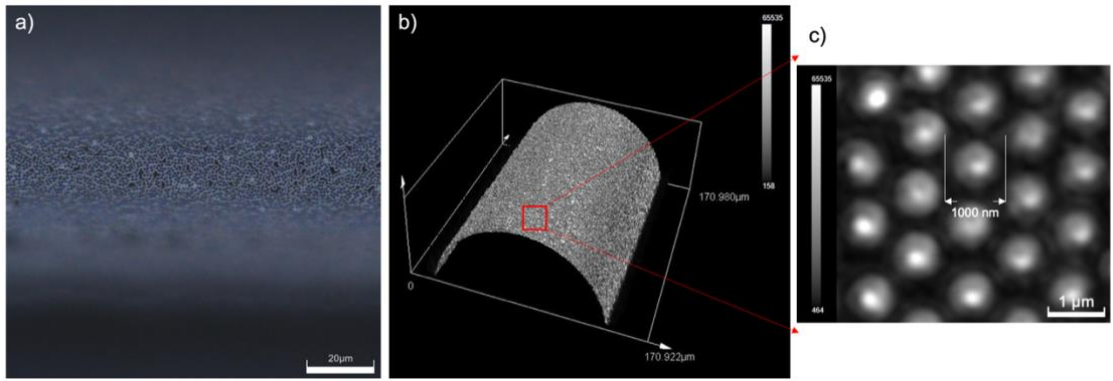

Fig. 3. Microscopic images of the microsphere-coated LPG surface. a) Dark field optical image, b) 3D confocal image of the curved surface coated with microspheres, and c) Magnified view of the microsphere array from (b).

### 3.2 Morphology of nanostructured LPG

Figure 4 illustrates the surface morphologies of LPG1 and LPG2 samples, following superlens-assisted laser nanostructuring for single-time and double-time, respectively. Nanoholes are clearly visible in dark-field optical microscopy (Fig. 4a). Zoomed SEM (Fig. 4b) and AFM images (Fig. 4c-d) reveal that these nanoholes are round in shape, with sizes ranging from 300 to 500 nm and depths ranging from 60 to 120 nm. Variations in size and defects in hexagonal patterns may be attributed to imperfections in the microsphere array deposition quality before laser processing, non-flat-top Gaussian laser beams, and fluctuations in laser output energy during laser processing.

Comparing the LPG0 sample with the LPG1 sample, we observe that the double-time process results in higher nanohole densities. The calculated nanohole density, defined as the generated nanohole area divided by the total surface area of the fiber, is approximately 5% for the LPG0 sample and 7.9% for the LPG2 sample. Notably, the second processing (LPG1) was performed on a surface already possessing nanoholes from the initial fabrication process (LPG2), leading to somewhat uneven distribution and overlapping nanoholes (see LPG2 in Fig.4c). However, this overlapping does not introduce any adverse effects beyond increasing the overall nanohole density. While it is possible to repeat the process multiple times to further increase nanohole density, such extensions are beyond the scope of this study.

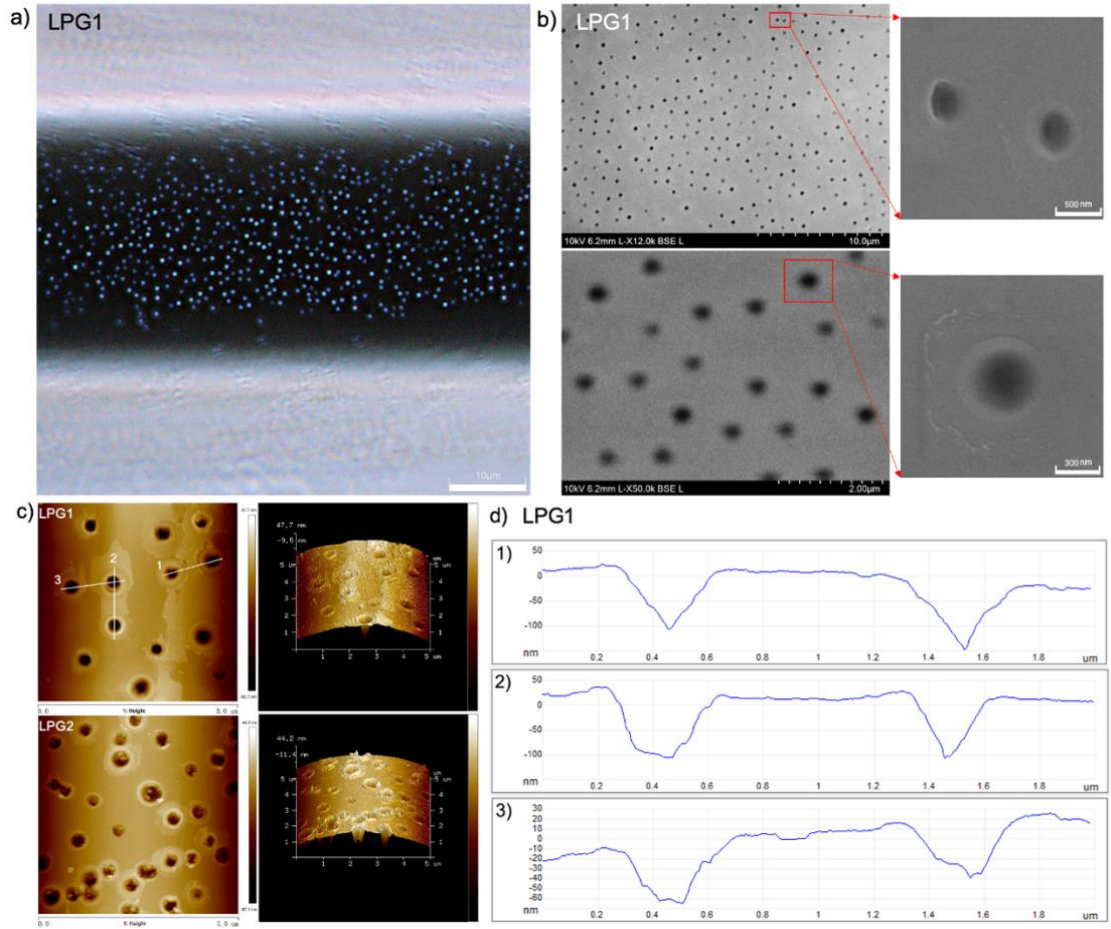

Fig. 4. Surface morphologies of processed LPG1 and LPG2 samples. a) Dark-field optical image of LPG1 sample. b) SEM image LPG1 sample, c) AFM images of LPG1 and LPG2 samples, d) depth profile of the LPG1 sample in c.

*3.3 RI sensitivity Measurement*

In a series of experiments, LPG1 with a 5% nanohole density resulting from one-time nanostructuring was exposed to sucrose aqueous solutions ranging from 0% (DI water) to 70%. Figure 6a shows a significant 480 pm red shift under ambient air conditions after nanohole structuring. The total peak shift increased from 7.05 nm (unprocessed) to 8.16 nm, marking a 15.7% total shift increase (Figure 6b-c). For LPG2, featuring double-time nanostructuring with a 7.9% nanohole density, experiments involved glycerin aqueous solutions ranging from 0% to 90%. Figure 6d demonstrates a prominent 530 pm red shift under air conditions following nanohole structuring. The total peak shift increased from 8.79 nm (unprocessed) to 10.70 nm, reflecting a 21.7% total shift increase (Figure 6e-f).

The peak red-shift effect results from generated nanoholes on the fiber, leading to a reduced effective thickness of the cladding layer but increased surface contact area with the sensing medium [39,40]. The physics of the red shift will be discussed in the modeling and simulation section.

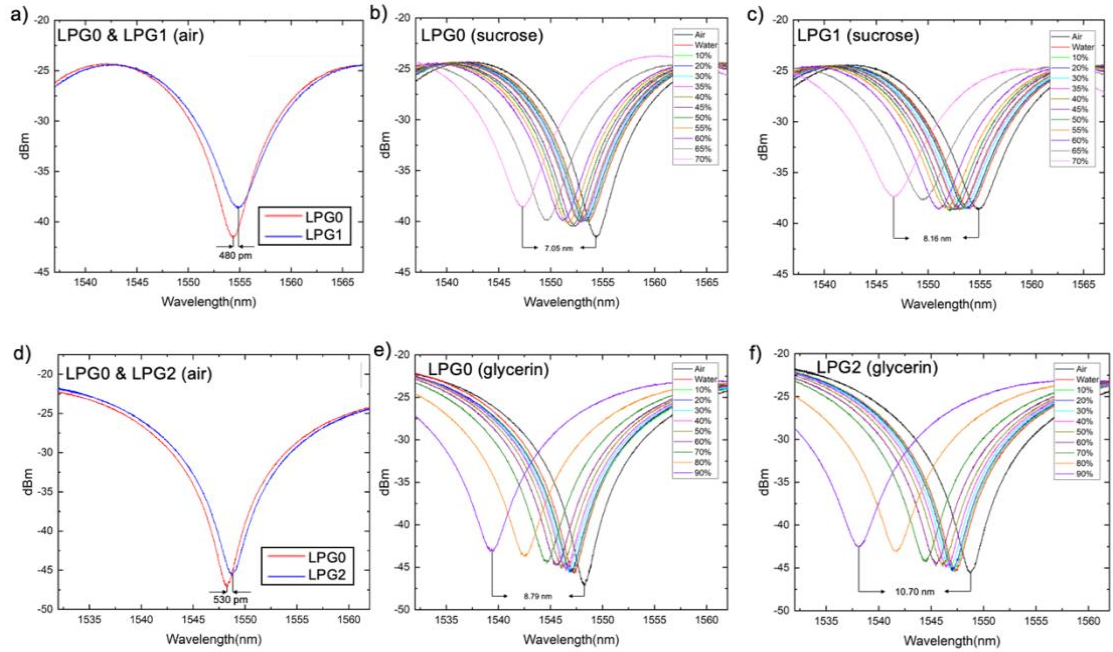

*Fig. 6. Transmission spectrum of (a) LPG1 in air, before and after nanostructuring, (b-c) LPG1 sensing of sucrose solution at various concentrations before and after nanostructuring, (d) LPG2 in air, before and after nanostructuring, (e-f) LPG2 sensing of glycerin solution at various concentrations before and after nanostructuring.*

The relationships between solution weight ratio and refractive index (RI) for both sucrose and glycerin aqueous solutions are shown in Figure 7a, displaying linear characteristics as observed in [36]. Figure 7b presents the shifted peak locations of two LPGs, indicating their response to different solution concentrations before and after nanohole structuring. The slopes of these curves represent RI sensitivity, a critical sensing parameter that improves as the surrounding RI approaches the cladding material's RI [37].

The results clearly demonstrate that nanostructured LPG1 and LPG2 samples exhibit significantly higher sensitivity compared to their original counterparts, as depicted in Fig. 8(a) and Fig. 8(b), respectively. LPG1 shows an average sensitivity increase of 16.08%, while LPG2 demonstrates an even more substantial average sensitivity increase of 19.57%. These experiments underscore the superior performance of nanostructured LPGs over the original unprocessed LPGs, with enhancements particularly pronounced in higher nanohole density structures. The increased total peak shifts indicate the improved performance of nanostructured LPGs and their potential for various optical and sensing applications. Ongoing efforts are directed toward refining methods to achieve even higher nanohole density samples, further advancing the sensitivity of nanostructured LPGs in refractive index-based applications.

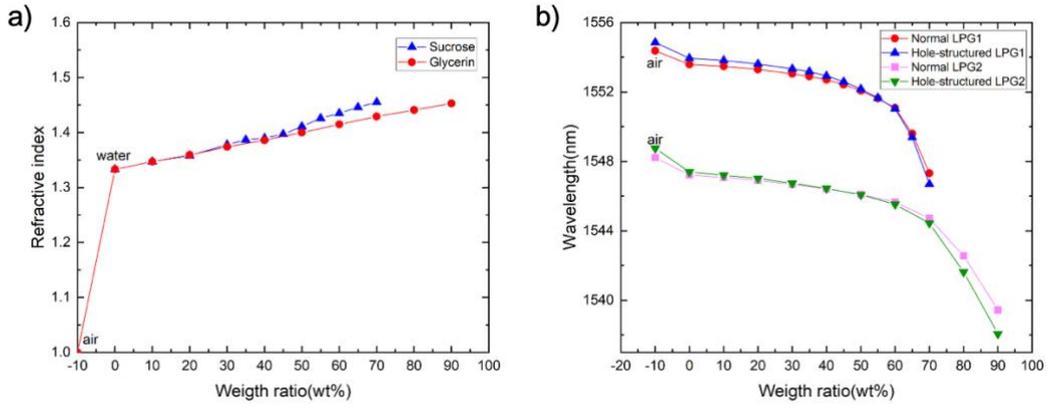

Fig. 7. (a) Relationships between solution weight ratio and refractive index, (b) shifted peak locations of two LPGs at varying solution concentrations.

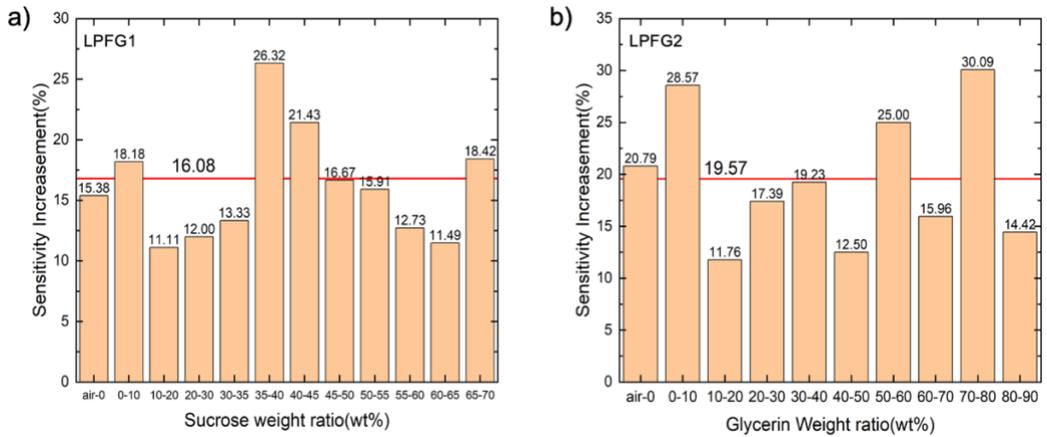

Fig. 8. Sensitivity increase of (a) LPG1 for sucrose solution concentration interval, (b) LPG2 for glycerin solution concentration interval.

*3.4 Modelling and Simulation*

To better understand the enhanced sensitivity of nanostructured LPGs, we developed a MATLAB model rooted in optical fiber waveguides coupled mode theory [35-37]. This model computes LPG transmission spectra and investigates the correlation between peak shifts and variations in the surrounding refractive index (RI). A key feature of our model is its consideration of the effective reduced thickness of the cladding layer to account for the influence of generated nanoholes on the LPGs. Our model parameters closely mirror those of the experimental LPG, featuring a cladding diameter of 125 μm, core diameter of 8.2 μm, grating length of 20 mm, and a grating period of 550 μm. We specifically focus on the highest-order peak, renowned for its heightened sensitivity to RI changes [38].

Figures 5 (a,d) depict the peak shifts for the original, unprocessed LPG0 sample. The total peak shift spans 8.04 nm as the RI varies from 1.0 to 1.445, serving as the baseline for subsequent comparisons with processed samples. In comparison, for the LPG1 sample subjected to a single-time nanostructuring process, the effective cladding layer thickness is reduced by 6 nm through depth reduction calculations due to the presence of generated nanoholes (5% nanohole density). This leads to a 600-pm red-shift of the transmission peaks in an air environment when compared to the LPG0 sample under the same conditions. Over the RI range from 1.0 to 1.445, the cumulative shift amounts to 8.34 nm for the LPG1 sample, representing a 3.7% increase compared to the unprocessed LPG0 sample, as shown in Figs 5 (b,e). Similarly, for the LPG2 sample with double-time nanostructuring, a larger equivalent reduction in cladding layer thickness is anticipated, now amounting to 9.5 nm. As depicted in Figs 5 (c,f), this results in a 970-pm red-shift of the peak wavelength in an air environment. Over the entire RI range, the total shift reaches 8.38 nm, reflecting a 4.3% increase compared to the unprocessed LPG.

Our model provides a qualitative explanation for the primary effects of nanostructured holes on LPG sensing performance: the peak red-shift effect (theoretical 600–970 pm redshift versus experimental 480–530 pm redshift) and the enhanced sensitivity effect (theoretical 3.7–4.3% vs experimental 16.08–19.57%). The quantitative discrepancy between theory and experiment indicates the presence of additional physics that our current simplified model does not capture. For instance, factors such as the shapes of the nanoholes and their increased surface contact area with the surrounding medium may play a significant role. Our next step involves developing a more precise model using modern FEM software, such as COMSOL, to further explore these complexities.

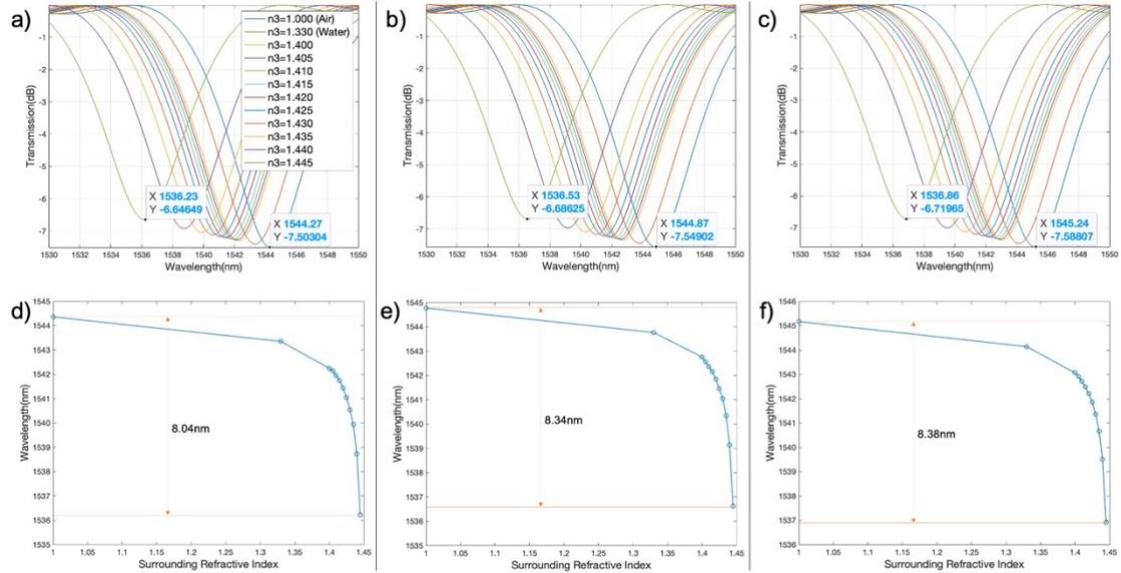

Fig. 5. Simulated transmission curves and corresponding peak wavelength as function of RI ranging from 1.0 to 1.445 for (a,d) LPG0 (original), (b,e) LPG1 (5% hole density) and (c,f) LPG2 (7.9% hole density) samples, respectively.

## 4. Conclusion

In conclusion, our study presents an innovative approach to enhance the sensitivity of Long Period Optical Fiber Gratings (LPGs) for refractive index sensing. We achieved this enhancement by creating nanohole-structured LPGs (NS-LPGs) through a novel microsphere-assisted superlens laser nanostructuring technique. Our experiments compared NS-LPGs with different nanohole densities in sucrose and glycerin solutions. The results demonstrated a remarkable increase in sensitivity, with NS-LPGs showing enhancements of 16.08% and 19.57% compared to traditional LPGs. Higher nanohole density resulted in greater sensitivity improvements. Importantly, these nanohole structures offer durability in challenging environments, addressing common issues associated with conventional surface-coated LPGs. Our research represents a significant advancement in LPG sensor technology, emphasizing the potential of surface nanostructuring to enhance refractive index sensing capabilities. This innovation holds promise for a wide range of practical applications.